\documentclass[preprint,amssymb,amsmath,aps,floatfix,prl,nofootinbib]{revtex4-1}

\newcommand{\beq}{\begin{eqnarray}}
\newcommand{\eeq}{\end{eqnarray}}

\begin{document}

\title{Twist--three relations of gluonic correlators for the transversely polarized nucleon }

\author{Yoshitaka Hatta$^{1}$\footnote{E-mail: hatta@het.ph.tsukuba.ac.jp}}
\author{Kazuhiro Tanaka$^{2}$\footnote{E-mail: tanakak@sakura.juntendo.ac.jp}}
\author{Shinsuke Yoshida$^{1}$\footnote{E-mail:  yoshida@het.ph.tsukuba.ac.jp}}
\affiliation{${}^1$\small{Faculty  of Pure and Applied Sciences, University
of Tsukuba, Tsukuba, Ibaraki 305-8571, Japan}\\
${}^2$\small{Department of Physics, Juntendo University, Inzai, Chiba 270-1695, Japan}}

\begin{abstract}
We derive exact relations among the polarized gluon and three--gluon distributions in the transversely polarized nucleon which are relevant to single and double spin asymmetries in various hard processes. We also discuss the partonic decomposition of the transverse nucleon spin and point out a potential problem with frame--independence.

\end{abstract}

\maketitle

\section{1. Introduction}

%Spin structure functions in transversely polarized nucleon are going to be determined
%through various types of hard processes in ongoing and future experiments.

Twist--three effects in QCD spin physics have been a subject of much interest over the past  decade or so, primarily due to their relevance to the single--spin asymmetry (SSA). In the collinear factorization framework, the SSA is a unique observable in QCD in which the first nonvanishing contribution comes solely from the `genuinely twist--three', multi-parton correlations in the transversely polarized nucleon. Notwithstanding the presumed suppression by a hard scale, sizable SSAs have been observed in light hadron ($h$) productions
in the semi-inclusive deep inelastic scattering (DIS)
$ep^\uparrow \rightarrow ehX$
and in the $pp$--collision $pp^\uparrow \rightarrow hX$~\cite{Barone:2010zz}.  A related observable, the transverse--longitudinal double--spin  asymmetry $A_{LT}$ in
Drell--Yan experiments, gives access to the polarized quark distribution $g_T(x)$ \cite{Jaffe:1991ra,Koike:2008du}
which is also twist--three. Its first moment is the quark helicity contribution to the transverse spin $ \int dx\, g_T(x)=\Delta q$ which actually coincides with the usual quark polarization in the longitudinal spin.
A closer analysis based on the
%local~\cite{Bukhvostov:1984as,Ji:1990br,Kodaira:1996md}
%and nonlocal~\cite{Balitsky:1987bk,Eguchi:2006qz}
operator product expansion (OPE)~\cite{Bukhvostov:1984as,Ji:1990br,Kodaira:1996md,Balitsky:1987bk,Eguchi:2006qz} reveals that $g_T(x)$ is actually a quantity beyond a (polarized) parton density and embodies the genuinely twist--three, quark--gluon correlations inside the nucleon.

%It appeared as $G_3(x)$~\cite{Ji:1992eu},
In this paper, we discuss the polarized gluon distribution in the transversely polarized nucleon ${\cal G}_{3T}(x)$~\cite{Ji:1992eu,Kodaira:1998jn} (defined in Eq.~(\ref{1}) below\footnote{The same function was denoted by  $G_3(x)$ in ~\cite{Ji:1992eu}.}) which  is the gluonic counterpart of  $g_T(x)$, having the same spin, twist, charge--conjugation--even, and chiral--even properties. As in the case of $g_T$, ${\cal G}_{3T}$ consists of multi-parton components, and we will investigate its precise twist structure using the nonlocal version of the OPE~\cite{Balitsky:1987bk,Eguchi:2006qz}. The result of this is a set of exact identities which relate ${\cal G}_{3T}$ to a certain integral of  three--gluon light--cone correlation functions. Not only do these three--gluon correlators emerge in the QCD evolution of $g_T(x)$
%(or $g_2(x,Q^2)$ ~\cite{Bukhvostov:1984as,Mueller:1997yk,Kodaira:1997ig,Braun:2000yi})
and the quark--gluon correlators~\cite{Bukhvostov:1984as,Mueller:1997yk,Kodaira:1997ig,Braun:2000yi,Braun:2009mi}, but already at leading order they  contribute to  SSAs in pion (light hadron) production, jet production, Drell--Yan and direct photon production~\cite{Ji:1992eu,Koike:2011nx} at high $P_T$, and also to $A_{LT}$ with various high-$P_T$ final states~\cite{Metz:2012fq}. As a matter of fact, there is \emph{a priori} no fundamental argument that the three--gluon contribution is negligible compared to the quark--gluon contribution in these processes. Moreover, in high-$P_T$ open charm productions $pp^\uparrow \rightarrow DX$~\cite{Kang:2008ih,Koike:2011mb}
the three--gluon contribution is expected to even dominate over the quark--gluon contribution, and in $ep^\uparrow \rightarrow eDX$
only the three--gluon distribution contributes to the SSA~\cite{Kang:2008qh,Beppu:2010qn,Koike:2011ns,Beppu:2012vi}.

%Also, the quark--gluon correlation functions contribute to
%the SSA in  the Drell--Yan and direct photon productions~\cite{Koike:2011nx},
%and for the double--spin longitudinal--transverse asymmetry $A_{LT}$ in the similar processes (see e.g. \cite{Metz:2012fq}).
%By a straitforward manipulaiton in the light-cone quantization formalism,
%we recognize that $G^{\perp}(x)$ as well as $g_T(x)$
%is a quantity beyond (polarized) parton densities and
%embodies multiparton correlations
%inside the nucleon~\cite{Ji:1992eu,Kodaira:1998jn}.
%For $g_T(x)$, explicit form of the corresponding twist-3 relation is also obtained
%using the operator product expansion in the local~\cite{Ji:1990br,Kodaira:1996md}
%and nonlocal version~\cite{Balitsky:1987bk,Eguchi:2006qz},
%where the fields are not restricted on
%the light-cone and thus the constraints from Lorentz invariance are fully taken into account.

%The QCD radiative corrections mix
%those three-gluon light-cone correlation functions into the twist-3 quark distribution $g_T(x)$ and
%the associated quark-gluon correlation functions,
%and this mixing embodies, e.g., in the transvese-spin structure function $g_2(x, Q^2)$
%for the inclusive deep inelastic scattering,
%in the longitudinal-transverse asymmetry $A_{LT}$ for the polarized Drell-Yan process (see %e.g., \cite{Jaffe:1991ra,Koike:2008du}),
%and in the SSA for various processes~\cite{Braun:2009mi}.

Another issue we would like to clarify is the relation between the two types of three--gluon correlators. The above--mentioned `genuinely twist--three' correlator is defined in terms of the matrix element
$\sim \langle P S_\perp| F^{+\perp}F^{+\perp}F^{+\perp} | P S_\perp\rangle$
 with the gluon field--strength tensor $F^{\mu\nu}$ on the light-cone~\cite{Beppu:2010qn}.
  In addition to this `F--type' correlator, one may define the  `D--type' correlator  $\sim \langle P S_\perp| F^{+\perp}D^{\perp}F^{+\perp} | P S_\perp\rangle$,
with $D^\perp$ being the transverse component of the covariant derivative, which contains the twist--two as well as twist--three contributions.
%so--called Wandzura--Wilczek part.
These two types of correlators are actually not independent,  and problems may arise when trying to relate observables computed in different gauges and expressed in different types of correlators. For instance, both types of correlators appear in the cross section formula for SSA in Ref.~\cite{Ji:1992eu} which works in the light--cone gauge, whereas only the F--type correlator appear  in Refs.~\cite{Beppu:2010qn,Koike:2011mb,Koike:2011nx} working in the Feynman gauge. Our result  will help establish the equivalence of such different--looking expressions. The desired relation can be derived in an analogous way as in the corresponding relation
between the F--type and D--type quark--gluon correlators~\cite{Eguchi:2006qz} relevant to $g_T(x)$.

%in the cross section formulas~\cite{Ji:1992eu}.
%In some literature,
%only the D-type correlator arises in the cross section formulas~\cite{Belitsky:%2000pb},
%or both the F-type and D-type correlators arise simultaneously
%in the cross section formulas~\cite{Ji:1992eu}.
%We will derive the explicit relation to reexpress the D-type correlator using the F-type correlator.
%demonstrate that the F-type and D-type correlators are not all independent.
%The idea is to compare the two formlas for ${\cal G}_{3T}(x)$,
%one is obtained in the light-cone formalism
%and another is the above-mentioned new result based on the nonlocal version of the operator
%product expansion.

 Finally, as a byproduct of our analysis of ${\cal G}_{3T}$, we shall discuss the decomposition of the transversely polarized nucleon spin. It is indeed relevant and timely to do so in view of the recent surge of interest in this problem (mostly in the longitudinally polarized case, but also in the transversely polarized case; see Ref.~\cite{Lorce:2012rr} and references therein). Just as  the integral of $g_T(x)$ gives the quark helicity $\Delta q$,  we show that the first moment  of the twist--three distribution ${\cal G}_{3T}(x)$ is equal to the usual gluon helicity contribution $\Delta G$ to the nucleon spin. We will further discuss the implication of this result
%for the decomposition of the transverse spin
through an analysis based on the Pauli--Lubansky vector \cite{Harindranath:1999ve,Ji:2012vj}
 and the newly proposed decomposition scheme of the nucleon spin \cite{Chen:2008ag,Wakamatsu:2010cb,Hatta:2011zs}.

%As is well known, the operator-product-expansion analysis
%shows that

\section{2. Twist--two and twist--three gluon correlators}

\subsection{2.1. Two--gluon correlator}
Let us start with the two--gluon correlator.
 Up to  twist--three, and using the light--cone coordinates $x^{\pm}=x_{\mp}=\frac{1}{\sqrt{2}}(x^0\pm x^3)$, we define the matrix element of the two--gluon operator in a polarized nucleon state as
\beq
&&\int \frac{d\lambda}{2\pi} e^{i\lambda x} \langle PS| F^{+\alpha}(0) W_{0\lambda}F^{+\beta}(\lambda n) |PS\rangle
=-\frac{1}{2}xG(x) (P^+)^2 (g^{\alpha\beta}-P^\alpha n^\beta -P^\beta n^\alpha)
 \nonumber \\
 && \qquad \qquad \qquad \qquad \qquad -\frac{i}{2}x\Delta G(x) P^+ \epsilon^{+-\alpha\beta}S^+ -ix{\cal G}_{3T}(x)P^+ \epsilon^{+\alpha\beta \gamma}S_{\perp \gamma} + \cdots\,,  \label{1}
%&& =-\frac{1}{2}xG(x) (P^+)^2 (g^{\alpha\beta}-P^\alpha n^\beta -P^\beta n^\alpha)    \nonumber \\
%&& \qquad - ix\epsilon^{+\alpha\beta\mu}\left( \left(\frac{1}{2}\Delta G(x) -G_3(x)\right) S^+ P_\mu +G_3(x)P^+S_\mu  \right)
 \eeq
 where $n^\mu=\delta^\mu_- /P^+$ is a lightlike vector. We  use Greek letters for four--vector indices $\mu,\nu=\pm, 1,2$, and Latin letters $i,j=1,2$ for the transverse coordinates. We shall also use the two--dimensional antisymmetric tensor $\epsilon^{ij}=\epsilon^{-+ij}$, $\epsilon^{12}=+1$.   $P^\mu$ and $S^\mu$ are the momentum and spin vectors, respectively, normalized as $P^2=-S^2=M^2$.   $W_{0\lambda}$ is the Wilson line along the light--cone which makes nonlocal operators gauge invariant.  The unpolarized/polarized gluon distributions $G(x)$/$\Delta G(x)$ are standard~\cite{Ji:1992eu,Kodaira:1998jn}, whereas our focus in this paper is ${\cal G}_{3T}(x)$ which is relevant to the transverse polarization $S_\perp^\mu \equiv \delta^\mu_i S^i$.

%  or equivalently,
%\beq
%\epsilon_{+\alpha\beta\mu} \int \frac{d\lambda}{2\pi} e^{i\lambda x} \langle PS| F^{+\alpha}(0) WF^{+\beta}(\lambda n) |PS\rangle
% \sim 2 ix\left( \left(\frac{1}{2}\Delta G(x) -G_3(x)\right) S^+ P_\mu +G_3(x)P^+S_\mu  \right) \label{n}
% \eeq

From (\ref{1}), one can derive the following compact expression
\beq
\int \frac{d\lambda}{2\pi} e^{i\lambda x}\langle PS|F^{+\alpha}(0)W\tilde{F}^{\mu}_{\ \ \alpha}(\lambda n)|PS\rangle = ix\Delta G(x) S^+ P^\mu + 2ix{\cal G}_{3T}(x) P^+ S^\mu_\perp \nonumber \\
=
ix\bigl\{  (\Delta G(x)-2{\cal G}_{3T}(x)) S^+ P^\mu +2{\cal G}_{3T}(x)P^+S^\mu \bigr\} \,, \label{ret}
\eeq
 which is valid to twist--three accuracy, that is, ignoring the twist--four component corresponding to $\mu=-$.  [Hereafter the subscripts on the Wilson line indicating the initial and final points will be omitted.] (\ref{ret}) is the gluonic counterpart to the polarized quark distributions~\cite{Jaffe:1991ra}
%We also note the following identity
%\beq
%F^{+\alpha}W\tilde{F}^{\mu}_{\ \ \alpha} = F^{\mu\alpha}W\tilde{F}^+_{\ \ \alpha} + \epsilon^{+\mu\alpha\beta}F_\alpha^{\ \gamma} WF_{\beta \gamma} \,. \label{on}
%\eeq
\beq
P^+\int \frac{d\lambda}{2\pi} e^{i\lambda x} \langle PS| \bar{\psi}(0)\gamma_5\gamma^\mu W\psi(\lambda n)|PS\rangle = 2 \bigl(g_1(x) S^+P^\mu + g_T(x) P^+S^\mu_\perp \bigr)\, ,
\label{gt}
\eeq
where $\gamma_5 \equiv -i \gamma^0 \gamma^1 \gamma^2 \gamma^3$.
It is well known~\cite{Kodaira:1998jn}
that $g_T(x)$ can be written as the sum of terms related to twist--two distributions (`Wandzura--Wilczek part') and genuine twist--three contributions. One of our goals in this paper is to perform a similar decomposition for ${\cal G}_{3T}(x)$.

  Let us quickly obtain the Wandzura--Wilczek part of ${\cal G}_{3T}(x)$, deferring a more detailed analysis to the next section. Take the $(n-1)$-th moment of (\ref{ret}):
\beq
&& \langle PS|F^{+\alpha}(iD^+)^{n-1}\tilde{F}^{\mu}_{\ \ \alpha}|PS\rangle
\nonumber \\
&& \qquad =i(P^+)^{n-1}\int_{-1}^1 dx\, x^n \big\{ (\Delta G(x)-2{\cal G}_{3T}(x)) S^+ P^\mu +2{\cal G}_{3T}(x)P^+S^\mu \bigr\}\,. \label{g1}
\eeq
The composite operator on the left--hand--side can be decomposed into those with a definite twist ($= {\rm dimension}-{\rm spin}$),
\beq
&& F^{+\alpha}(iD^+)^{n-1}\tilde{F}^{\mu}_{\ \ \alpha} \nonumber \\
&& = \frac{1}{n+1}\left(F^{+\alpha}(iD^+)^{n-1}\tilde{F}^{\mu}_{\ \ \alpha} +\sum^{n-1}_{k=1} F^{+\alpha} (iD^+)^{k-1} iD^\mu  (iD^+)^{n-k-1} \tilde{F}^{+}_{\ \ \alpha} + F^{\mu\alpha}(iD^+)^{n-1}\tilde{F}^{+}_{\ \ \alpha} \right) \nonumber \\
 && +\frac{1}{n+1} \left(nF^{+\alpha}(iD^+)^{n-1}\tilde{F}^{\mu}_{\ \ \alpha} -\sum^{n-1}_{k=1} F^{+\alpha} (iD^+)^{k-1} iD^\mu  (iD^+)^{n-k-1}  \tilde{F}^{+}_{\ \ \alpha} - F^{\mu\alpha}(iD^+)^{n-1}\tilde{F}^{+}_{\ \ \alpha} \right)\,. \nonumber
\eeq
Parameterizing the matrix elements of the first line (twist--two operator) and the second line (twist--three operator)
in terms of the corresponding reduced matrix elements $a_n$ and $d_n$, respectively,
as
\beq
\langle PS| F^{+\alpha}(iD^+)^{n-1}\tilde{F}^{\mu}_{\ \ \alpha} |PS\rangle
&=& \frac{2ia_n}{n+1} \left(S^\mu (P^+)^n + nS^+P^\mu (P^+)^{n-1} \right) \nonumber \\
 && + \frac{ 2ind_n}{n+1} (S^\mu P^+ -S^+P^\mu)(P^+)^{n-1}\,,
\eeq
and comparing with (\ref{g1}), we find
%\beq
%\frac{a_n + nd_n}{n+1} =\int dx\, x^n {\cal G}_{3T}(x) \qquad \frac{n(a_n-d_n)}{n+1} = \int dx\, x^n
%\left(\frac{1}{2}\Delta G(x) -{\cal G}_{3T}(x) \right)\,,
%\eeq
 \beq
 \frac{1}{2}\int_{-1}^1 dx\, x^n \Delta G(x) =a_n\,,  \qquad  \int_{-1}^1 dx\, x^n {\cal G}_{3T}(x)  =  \frac{a_n +nd_n}{n+1}\,.
 \eeq
This leads to
\beq
\int dx\, {\cal G}_{3T}(x)=\frac{1}{2} \int dx\, \Delta G(x) =a_0= \Delta G\,, \label{delta}
\eeq
where $\Delta G$ is the gluon polarization. Moreover, ${\cal G}_{3T}(x)$ can be written as ($x \ge 0$)
 \beq
 {\cal G}_{3T}(x) = \frac{1}{2} \int^1_x \frac{dz}{z} \Delta G(z) +\delta {\cal G}(x)\,. \label{gperp}
 \eeq
 The first term is the  Wandzura--Wilczek part and $\delta {\cal G}(x)$ is the genuinely twist--three contribution which integrates to zero: $\int dx \, \delta {\cal G}(x)=0$.

\subsection{2.2.  F--type three--gluon correlator }
Next we define the `F--type' twist--three gluon correlator \cite{Ji:1992eu,Beppu:2010qn}
\beq
&&\frac{1}{(P^+)^2}\int \frac{d\lambda}{2\pi}\frac{d\mu}{2\pi} e^{i\lambda x_1+i\mu(x_2-x_1)}\langle PS|F^{+i}(0)WgF^{+j}(\mu n) W F^{+k}(\lambda n)|PS\rangle \nonumber \\
&&= \Bigl(F(x_1,x_2)g^{ik}\epsilon^{+j \rho\sigma} - F(x_2,x_2-x_1)g^{ij}\epsilon^{+k \rho\sigma} -F(x_1,x_1-x_2)g^{jk}\epsilon^{+i \rho\sigma} \Bigr)S_\rho^\perp P_\sigma\,.
\label{fty}
\eeq
Here the color indices are contracted by the totally antisymmetric structure constants: $F^{+i}F^{+j}F^{+k}\equiv F^{+i}_a F^{+j}_b(T^b)_{ac} F^{+k}_c = if_{abc}F^{+i}_aF^{+j}_bF^{+k}_c$. The function $F(x_1,x_2)$ at special values of its arguments $F(x,x)$  and $F(x,0)$ is related to the gluonic contribution to
 the single--spin asymmetry (SSA) \cite{Beppu:2010qn}. [Note that $F(x_1,x_2)/2$ equals
the corresponding correlation function $N(x_1, x_2)$ in the notation of \cite{Beppu:2010qn}.]
The three--gluon correlator whose color indices are contracted by the totally symmetric  $d$--symbol also contributes to SSA, but it will  not be discussed here because it
is associated with a $C$--odd operator and bears no relation to ${\cal G}_{3T}$ and the D--type distribution defined below
which are $C$--even.
 By $PT$ invariance, $F$ satisfies
\beq
F(x_1,x_2)=F(x_2,x_1)\,.
\eeq
On the other hand, from symmetry under permutation of the gluon field strength tensor $F^{\mu\nu}$, we find
\beq
F(x_1,x_2)=-F(-x_1,-x_2)\,.
\eeq

We shall also need the F--type quark--gluon operator
\beq
\int \frac{d\lambda}{2\pi} \frac{d\mu}{2\pi} e^{i\lambda x_1+i\mu(x_2-x_1)}
\langle PS |\bar{\psi}(0) \gamma^+W gF^{+\alpha}(\mu n) W \psi(\lambda n)|PS\rangle \nonumber \\
 = P^+ \epsilon^{+\alpha\rho\sigma}S_\rho P_\sigma G_F(x_1,x_2)\,,  \label{fq}
\eeq
 with the symmetry property $G_F(x_1,x_2)=G_F(x_2,x_1)$. $G_F$ at special values of its arguments, e.g.,
$G_F(x,x)$ (`soft-gluon pole' \cite{Qiu:1991pp,Ji:2006vf,Kouvaris:2006zy,Eguchi:2006qz})
 and $G_F(x,0)$ (`soft-fermion pole' \cite{Efremov:1984ip,Koike:2007dg,Koike:2009ge}), contributes to the SSA.

\subsection{2.3. D--type three--gluon correlator}

Finally we define the D--type three--gluon correlator
\beq
&& \frac{1}{P^+}\int \frac{d\lambda}{2\pi}\frac{d\mu}{2\pi} e^{i\lambda x_1+i(x_2-x_1)\mu} \langle PS|F^{+i}(0)WD^{j}(\mu n) W F^{+k}(\lambda n)|PS\rangle \nonumber \\
&&=\left(D_1(x_1,x_2)g^{ik}\epsilon^{+j\rho\sigma}+D_2(x_1,x_2)g^{ij} \epsilon^{+k\rho\sigma} -D_2(x_2,x_1)g^{jk}\epsilon^{+i\rho\sigma}  \right)S_\rho P_\sigma\,.
\label{ddd}
\eeq
Unlike the F--type correlator~(\ref{fty}), we need two independent functions $D_1$ and $D_2$.\footnote{Ref.~\cite{Ji:1992eu} introduced four functions which later turned out to be redundant, see the discussion
in \cite{Belitsky:2000pb,Beppu:2010qn}.}
Their symmetry properties are:
\beq
D_1(x_1,x_2) = -D_1(x_2,x_1)\,.
\eeq
\beq
D_1(x_1,x_2)= D_1(-x_1,-x_2)\,, \qquad D_2(x_1,x_2)=D_2(-x_1,-x_2)\,.
\eeq
In Ref.~\cite{Ji:1992eu}, both the F--type and D--type correlators appear in the cross section formula for SSA computed
in the light--cone gauge. On the other hand, Refs.~\cite{Beppu:2010qn,Koike:2011mb,Koike:2011nx} worked in the Feynman gauge and
 obtained a formula which contains only the F--type correlator, as a
result of a rather sophisticated reorganization of
the collinear expansion of the Feynman amplitudes using Ward identities.
Thus, even in a covariant gauge, one may achieve an expression of the cross section in terms of the D--type correlators depending on the intermediate steps of the calculations.\footnote{The
NLO correction to the deep-inelastic-scattering structure function $g_2$ was obtained in terms of the D--type correlators in \cite{Belitsky:2000pb},
while the same contribution was calculated in terms of the F--type correlators in \cite{Braun:2000yi}.}
 In order to relate these seemingly different results, one has to establish relations between the D--type and F--type distributions, to which we now turn. Note that, because the SSA is a `genuinely twist--three' effect
as demonstrated in \cite{Beppu:2010qn,Koike:2011mb,Koike:2011nx,Eguchi:2006mc}, the
%Wandzura--Wilczek
twist--two parts of the D--type function (see below) must cancel in the cross section formula.

\section{3. Relations between  gluonic correlators}

 The various quark and gluon correlators introduced in the previous section are not independent as they are related by the equations of motion. In this section we show that ${\cal G}_{3T}(x)$ and $D_{1,2}(x_1,x_2)$ can be entirely written by the other distributions $\Delta G(x)$, $F(x_1,x_2)$ and $G_F(x_1,x_2)$.

  Firstly, following \cite{Eguchi:2006qz} one can derive a general relation between the F--type and D--type correlators which in the present case reads
\beq
D_1(x_1,x_2) = {\mathcal P}\frac{F(x_1,x_2)}{x_1-x_2}\,,  \label{dd}
\eeq
\beq
D_2(x_1,x_2) = - {\mathcal P}\frac{F(x_2,x_2-x_1)}{x_1-x_2} + \delta(x_1-x_2)\tilde{g}(x_1)\,,
\label{del}
\eeq
 where ${\mathcal P}$ denotes the principal value and $\tilde{g}$ is defined via the following relation
 \beq
 && \int \frac{d\lambda}{2\pi}e^{i\lambda x} \langle PS|F^{+i}(0) W_{0\lambda} \left(D^j (\lambda n) +\frac{ig}{P^+} \int d\mu \frac{\epsilon(\mu-\lambda)}{2}W_{\lambda \mu} F^{+j}(\mu n) W_{\mu\lambda} \right)F^{+k}(\lambda n)  |PS\rangle \nonumber \\
 && \qquad = (P^+)^2 \epsilon^{ik}S_\perp^j \tilde{g}(x)\,.
\eeq
 We shall later see that $\tilde{g}$ is not an independent function,\footnote{We note that the operator in the parentheses on the left--hand--side
can be written as $D^j(\lambda n)- ig A_{phys}^j(\lambda n)\equiv D^j_{pure}(\lambda n)$
using (\ref{aphys}) below with ${\mathcal K}(\lambda)=\frac{1}{2}\epsilon(\lambda)$. $D_{pure}^\mu$ is the covariant derivative associated with the `pure gauge' part of the gluon field~\cite{Hatta:2011zs}.} but can be entirely expressed in terms of $\Delta G$, $G_F$, and $F$.

Next we derive a formula which relates ${\cal G}_{3T}(x)$ to the genuine twist--three correlators, namely, we determine the function $\delta {\cal G}(x)$ in (\ref{gperp}).
For this purpose, we employ the `nonlocal OPE' approach of Refs.~\cite{Balitsky:1987bk,Eguchi:2006qz} and consider  the following matrix element
\beq
I \equiv z^\alpha \left(\frac{\partial}{\partial z^\alpha}  z_\nu\langle PS|F^{\nu\gamma}(0)W\tilde{F}^{\mu}_{\ \ \gamma}(z)|PS\rangle
-(\alpha \leftrightarrow \mu) \right)\,.  \label{two}
\eeq
   In (\ref{two}), $z^\mu$ is generic, not necessarily proportional to $n^\mu$, and this ensures that the constraints from Lorentz invariance are fully taken into account. The covariant expansion of the nonlocal operators in powers of $z^2$
is equivalent to the OPE according to the deviation from the light-cone
(i.e., the twist expansion), and one can identify exact relations among operators belonging to the same twist.

Let us calculate (\ref{two}) in two ways. On one hand, we can directly evaluate $I$ in terms of the gluon distributions by generalizing   the definition (\ref{ret}) away from the light--cone
\beq
&& z_\nu\langle PS|F^{\nu\alpha}(0)W\tilde{F}^{\mu}_{\ \ \alpha}(z)|PS\rangle \nonumber \\
&& \quad = i\int dx\, e^{-ixP\cdot z}
 x\bigl\{  (\Delta G(x)-2{\cal G}_{3T}(x)) S\cdot z P^\mu +2{\cal G}_{3T}(x)P\cdot z S^\mu \bigr\}\,.
 \label{off}
\eeq
Plugging this into (\ref{two}), we find
\beq
I= iz^- \int dx\, e^{-ixP^+z^-}\left(2x^2\frac{\partial}{\partial x}{\cal G}_{3T}(x) +x\Delta G(x) \right) (P^\mu S^+ -S^\mu P^+)\,,  \label{gp}
\eeq
where we set $z^\alpha=\delta^\alpha_- z^-$ after differentiation.
 On the other hand, we can apply the $z$--derivative on fields and Wilson lines, and then use the equations of motion,
\beq
I&=& z^- \langle PS| \Bigl(F^{+\gamma}W\tilde{F}^\mu_{\ \, \gamma} -F^{\mu\gamma}W\tilde{F}^+_{\ \, \gamma} \Bigr)
 \nonumber \\
 && \quad +z^- F^{+\gamma}\left(WD^+\tilde{F}^\mu_{\ \, \gamma} -WD^\mu \tilde{F}^+_{\ \, \gamma} +iz^- \int_0^1 dt\ tWgF^{+\mu} (tz) W\tilde{F}^+_{\ \gamma}(z) \right)|PS\rangle\,, \label{w2}
\eeq
 where again we set $z^\alpha=\delta^\alpha_- z^-$ after differentiation. With the help of an identity
 \beq
\epsilon^{\alpha\beta\gamma\sigma}D^\lambda F_{\lambda \sigma} = D^\alpha \tilde{F}^{\beta\gamma} + D^\beta \tilde{F}^{\gamma\alpha} + D^\gamma \tilde{F}^{\alpha\beta}\,,
\label{for}
\eeq
 and the equations of motion, $D_\nu F^{\nu\mu}=g\bar{\psi}\gamma^\mu \psi$, where
the color matrix $T^a$ is suppressed, the second line  of (\ref{w2}) takes the form
\beq
&& (z^-)^2 \langle PS| F^{+\gamma}\left(\epsilon^{+\mu}_{\ \ \ \gamma\sigma}WD_\lambda F^{\lambda \sigma} -W D_\gamma \tilde{F}^{+\mu} +iz^-\int_0^1 dt\ tWgF^{+\mu} (tz) W\tilde{F}^+_{\ \gamma}(z)
\right) |PS\rangle \nonumber \\
&&  =(z^-)^2\langle PS|  \epsilon^{+\mu}_{\ \ \ \gamma\sigma} F^{+\gamma}(0)W g\bar{\psi}\gamma^\sigma \psi(z) -g\bar{\psi}\gamma^+\psi (0) W\tilde{F}^{+\mu}(z)
\nonumber \\
  &&  \qquad \qquad -iz^-\int_0^1 dt F^{+\gamma}WgF^+_{\ \, \gamma}W\tilde{F}^{+\mu} +iz^-\int_0^1 dt\ tF^{+\gamma}WgF^{+\mu}  W\tilde{F}^+_{\ \gamma}|PS\rangle\,.  \label{sec}
\eeq

As for the first line of (\ref{w2}), we use the following trick (see, e.g., \cite{Braun:2000yi})
\beq
&& F^{+\gamma}(0)W\tilde{F}^\mu_{\ \, \gamma}(z) -F^{\mu\gamma}(0)W\tilde{F}^+_{\ \, \gamma}(z) \nonumber \\
&&  \qquad = \int_0^1 du \frac{d}{du} \left( F^{+\gamma}(0)W\tilde{F}^\mu_{\ \, \gamma}(uz) -F^{\mu\gamma}(0)W\tilde{F}^+_{\ \, \gamma}(uz) \right)\,.  \label{trick}
\eeq
The contribution from the lower limit $u=0$ vanishes because of the following identity
\beq
F^{+\gamma}(z)W\tilde{F}^\mu_{\ \, \gamma}(z') -F^{\mu\gamma}(z)W\tilde{F}^+_{\ \, \gamma}(z')
=\epsilon^{+\mu\alpha\beta}F_\alpha^{\ \gamma}(z)W F_{\beta \gamma}(z')\,,
\eeq
 which gives zero when evaluated at equal points $z=z'$.
The matrix element of the right--hand--side of (\ref{trick}) will depend only on $uz^-$, so we can write
 $u\frac{d}{du}=z^-\frac{\partial}{\partial z^-}$ and the first line of (\ref{w2}) becomes
\beq
&& z^- \int_0^1 du \frac{z^-}{u}\frac{\partial}{\partial z^-} \langle PS| F^{+\gamma}(0)W\tilde{F}^\mu_{\ \, \gamma}(uz) -F^{\mu\gamma}(0)W\tilde{F}^+_{\ \, \gamma}(uz) |PS\rangle \nonumber \\
%&=&  (z^-)^2 \int_0^1 du  \langle PS| \biggl\{ F^{+\gamma}(0)W\bigl(-D^\mu\tilde{F}_\gamma^{\ +} -D_\gamma \tilde{F}^{+\mu}  + \epsilon^{+\mu}_{\ \ \ \gamma\delta}D_\lambda F^{\lambda \delta} \bigr)  -\left(F^{\gamma+}\overleftarrow{D}^\mu +F^{+\mu}\overleftarrow{D}^\gamma\right) W \tilde{F}^{+}_{\ \,\gamma} \biggr\} |PS\rangle  \nonumber \\
&& = (z^-)^2 \int_0^1 du  \langle PS| \biggl\{ F^{+\gamma}\overleftarrow{D}_\gamma W\tilde{F}^{+\mu} + F^{+\mu}WD^\gamma \tilde{F}^+_{\ \, \gamma}  +\epsilon^{+\mu}_{\ \ \ \gamma\delta}F^{+\gamma}WD_\lambda F^{\lambda \delta}
\nonumber \\
&& \qquad \qquad -iuz^- \int_0^1 dt \Bigl(-F^{+\gamma}(0)WgF^{+\mu}(tuz)W\tilde{F}^+_{\ \, \gamma}(uz) \nonumber \\
 && \qquad \qquad \qquad \qquad + F^{+\gamma}WgF^{+}_{\ \, \gamma} W \tilde{F}^{+\mu} + F^{+\mu}WgF^{+\gamma}W\tilde{F}^+_{\ \, \gamma}  \Bigr) \biggr\} |PS\rangle\,,   \label{mat}
\eeq
 where we again used (\ref{for}) and the Bianchi identity, $D^\alpha F^{\beta\gamma}+D^\beta F^{\gamma\alpha} + D^\gamma F^{\alpha\beta}=0$.

Comparing (\ref{gp}) with (\ref{sec}) and (\ref{mat}), and noting that $D_\gamma\tilde{F}^{\gamma+}=0$ from the Bianchi identity, we find
\beq
&& i \int dx\, e^{-ixP^+z^-}\left(2x^2\frac{\partial}{\partial x}{\cal G}_{3T}(x) +x\Delta G(x) \right) (P^\mu S^+ -S^\mu P^+) \nonumber \\
&&=z^-\langle PS|  \epsilon^{+\mu}_{\ \ \ \gamma\sigma} F^{+\gamma}(0)W g\bar{\psi}\gamma^\sigma \psi(z) -g\bar{\psi}\gamma^+\psi (0) W\tilde{F}^{+\mu}(z)
\nonumber \\
  &&  \qquad \qquad -iz^-\int_0^1 dt F^{+\gamma}WgF^+_{\ \, \gamma}W\tilde{F}^{+\mu} +iz^- \int_0^1 dt\ tF^{+\gamma}WgF^{+\mu}  W\tilde{F}^+_{\ \gamma}|PS\rangle
  \nonumber \\
&& \quad +  z^- \int_0^1 du  \langle PS| \biggl\{ -g\bar{\psi}\gamma^+\psi (0) W\tilde{F}^{+\mu}(uz)  +\epsilon^{+\mu}_{\ \ \ \gamma\delta}F^{+\gamma}(0)W g\bar{\psi}\gamma^\delta \psi(uz)
\nonumber \\
&& \qquad \qquad -iuz^- \int_0^1 dt \Bigl(-F^{+\gamma}(0)WgF^{+\mu}(tuz)W\tilde{F}^+_{\ \, \gamma}(uz)
 \nonumber \\
  && \qquad \qquad \qquad  + F^{+\gamma}WgF^{+}_{\ \, \gamma} W \tilde{F}^{+\mu} + F^{+\mu}WgF^{+\gamma}W\tilde{F}^+_{\ \, \gamma}  \Bigr) \biggr\} |PS\rangle\,.
\eeq

The matrix elements on the right--hand--side can be evaluated by the F--type  correlators $F(x_1,x_2)$  and $G_F(x_1,x_2)$ (cf., (\ref{fty}), (\ref{fq})).   After very tedious calculations, we get
\beq
&& 2x^2\frac{\partial}{\partial x}{\cal G}_{3T}(x) +x\Delta G(x) =\int dX \left(-2\frac{\partial}{\partial x} G_F(X,x) + \frac{2}{x}G_F(X,x) \right) \nonumber \\
&& \quad + \int dx' {\mathcal P}\frac{1}{x-x'} \Biggl[ \left(\frac{\partial}{\partial x} + \frac{\partial}{\partial x'}\right) \bigl(2F(x,x')-3F(x',x'-x)-3F(x,x-x') \bigr)  \nonumber \\ &&  \qquad \qquad -  \left(\frac{\partial}{\partial x} - \frac{\partial}{\partial x'}\right) \bigl(F(x',x'-x)-F(x,x-x') \bigr) \Biggr] \nonumber \\
 && \qquad \quad + 4\int dx' {\mathcal P}\frac{1}{x(x-x')} \bigl(F(x',x'-x) +F(x,x-x') \bigr)\,, \label{28}
\eeq
 where we used the symmetry relations (10) and (11) and switched to the notation $X\equiv \frac{x_1+x_2}{2}$ and $x \equiv x_1-x_2$ in the arguments of $G_F$. [Note that $G_F(X,x)=G_F(X,-x)$.]
(\ref{28}) can be solved for ${\cal G}_{3T}(x)$ with the boundary condition ${\cal G}_{3T}(\pm 1)=0$.
\beq
{\cal G}_{3T}(x) &=& \frac{1}{2}\int_x^{\epsilon(x)}\frac{dx'}{x'}\Delta G(x') + \int_x^{\epsilon(x)}
\frac{dx'}{x'^3}\int dX \left(x'\frac{\partial}{\partial x'} G_F(X,x') -G_F(X,x') \right)
\nonumber \\
 &&-\frac{1}{2} \int_x^{\epsilon(x)} dx'' \int dx' {\mathcal P}\frac{1}{x''^2(x''-x')}  \nonumber \\ &&  \qquad \times \Biggl[ \left(\frac{\partial}{\partial x''} + \frac{\partial}{\partial x'}\right) \bigl(2F(x'',x')-3F(x',x'-x'')-3F(x'',x''-x') \bigr)  \nonumber \\ &&  \qquad \qquad -  \left(\frac{\partial}{\partial x''} - \frac{\partial}{\partial x'}\right) \bigl(F(x',x'-x'')-F(x'',x''-x') \bigr) \Biggr] \nonumber \\
 && \qquad  -2\int_x^{\epsilon(x)} dx'' \int dx' {\mathcal P}\frac{1}{x''^3(x''-x')} \bigl(F(x',x'-x'') +F(x'',x''-x') \bigr)\,,  \label{solve}
\eeq
 where $\epsilon(x)=x/|x|$.  The first term is the Wandzura--Wilczek part which agrees with (\ref{gperp}). The rest is the genuinely twist--three function of $G_F$ and $F$ which was denoted by $\delta {\cal G}(x)$ in (\ref{gperp}).

Let us also show the moments of (\ref{solve}):
\beq
&& \int_{-1}^1 dx \, x^{n-1}{\cal G}_{3T}(x) \nonumber \\
&& = \frac{1}{2n} \int_{-1}^1 dx\, x^{n-1}\Delta G(x)
+ \frac{1}{n} \int_{-1}^1 dx\, x^{n-1}
\left(\frac{1}{x}\frac{\partial}{\partial x} G_F(X,x) -\frac{1}{x^2}G_F(X,x) \right) \nonumber \\ && \quad -\frac{1}{2n} \int_{-1}^1 dxdx'\,   {\mathcal P}\frac{x^{n-1}}{x(x-x')} \Biggl[ \left(\frac{\partial}{\partial x} + \frac{\partial}{\partial x'}\right) \bigl(2F(x,x')-3F(x',x'-x)-3F(x,x-x') \bigr)  \nonumber \\ && \qquad   \qquad \qquad -  \left(\frac{\partial}{\partial x} - \frac{\partial}{\partial x'}\right) \bigl(F(x',x'-x)-F(x,x-x') \bigr) \Biggr] \nonumber \\
&& \qquad \qquad \qquad \qquad  -\frac{2}{n}\int_{-1}^1 dx dx' {\mathcal P}\frac{x^{n-1} }{x^2(x-x')} \bigl(F(x',x'-x) +F(x,x-x') \bigr)\,.
\eeq
When $n=1$, the $G_F$ terms vanish and we get, after integration by parts,
\beq
 \int_{-1}^1 dx \, {\cal G}_{3T}(x)  = \Delta G  &+& \int dx dx'\, \biggl\{{\mathcal P}\frac{1}{x(x-x')^2} \bigl(F(x',x'-x) -F(x,x-x')\bigr) \nonumber \\
 && \qquad \qquad -{\mathcal P}\frac{1}{x^2(x-x')} \bigl(F(x,x')+F(x,x-x')\bigr) \biggr\}\,.
\eeq
By the change of variables and the symmetry property $F(x,x')=F(x',x)$, it is easy to check that the two terms in the curly brackets are equal and they both vanish:
\beq
 && \int dx dx' {\mathcal P}\frac{1}{x(x-x')^2} \bigl(F(x',x'-x) -F(x,x-x')\bigr)
 = \int dx dx' {\mathcal P}\frac{1}{xx'(x'-x)} F(x',x'-x) \nonumber \\
 && \qquad \qquad \qquad \qquad \qquad \qquad \qquad \qquad \qquad =  \int dx dx' {\mathcal P}\frac{1}{xx'(x'-x)} F(x',x) =0\,.
  \eeq
  We therefore recovered  (\ref{delta}).\\

%\subsection{3.2. Relation between $\tilde{g}$ and $F$}

Finally in this section, we determine the function $\tilde{g}(x)$ defined in (\ref{del}).
Recall the  definition~(\ref{1}),
\beq
\int \frac{d\lambda}{2\pi} e^{i\lambda x} \langle PS|F^{+i}(0) W F^{+-}(\lambda n) |PS\rangle= -ix{\cal G}_{3T}(x)P^+\epsilon^{ij}S_j\,.
\eeq
After integration by parts, the left--hand--side can be written as
\beq
&& \frac{i}{P^+x}\int \frac{d\lambda}{2\pi} e^{i\lambda x} \langle PS|F^{+i}(0)WD^+F^{+-}(\lambda n)|PS\rangle  \nonumber \\
&& =\frac{-i}{P^+x} \int \frac{d\lambda}{2\pi} e^{i\lambda x} \langle PS| F^{+i}(0) W \left(D_jF^{+j}(\lambda n) +g\bar{\psi}(\lambda n)\gamma^+ \psi(\lambda n) \right)|PS\rangle \nonumber \\
&&   = \frac{iP^+}{x} \left(\int dx'\bigl(D_1(x',x)+D_2(x',x)-2D_2(x,x')\bigr) + \int dX G_F(X,x) \right)\epsilon^{ij}S_j\,,
\eeq
where in the second line we used the equations of motion, $D_\nu F^{\mu\nu}=-g\bar{\psi}\gamma^\mu \psi$,
and in the third line we substituted (\ref{fq}) and (\ref{ddd}).
We thus find, using (\ref{dd}) and (\ref{del}),
\beq
\tilde{g}(x) &=&x^2{\cal G}_{3T}(x)+  \int dX G_F(X,x) \nonumber \\
 && +\int dx' {\mathcal P} \frac{1}{x'-x} \bigl(F(x',x) -F(x,x-x') -2F(x',x'-x) \bigr)\,.
 \label{ind}
\eeq
Eliminating ${\cal G}_{3T}$ via (\ref{solve}) and doing integrations by parts, we  arrive at
\beq
\tilde{g}(x)
 &=&\frac{x^2}{2}\int_x^{\epsilon(x)}\frac{dx'}{x'}\Delta G(x') + x^2\int_x^{\epsilon(x)}
\frac{dx'}{x'^3}\int dX G_F(X,x')
\nonumber \\
 && +x^2 \int_x^{\epsilon(x)} dx'' \int dx' \biggl\{{\mathcal P} \frac{-2}{x''^3(x''-x')} \bigl(F(x'',x')-F(x',x'-x'') \bigr)  \nonumber \\ &&  \qquad \qquad + {\mathcal P}  \frac{1}{x''^2(x''-x')^2}  \bigl(F(x',x'-x'')-F(x'',x''-x') \bigr) \biggr\}\,.
\eeq

\section{4. Gluon helicity contribution to the transverse spin}

 Due to the property (\ref{delta}), it is natural to identify
%${\cal G}_{3T}(x)$ as the gluon polarization density and
$\Delta G=\int {\cal G}_{3T}(x)$ as the total gluon helicity contribution to the transversely polarized nucleon spin. In this section we show that this expectation is consistent with the newly developed framework of spin decomposition in QCD \cite{Chen:2008ag,Hatta:2011zs}.

Recently there has been a lot of debate (and confusion) about the proper decomposition of the transversely polarized nucleon spin \cite{Bakker:2004ib,Leader:2011cr,Ji:2012sj,Ji:2012vj}. We first note that it is important to distinguish two `transverse spin operators' considered in the literature. One is the QCD angular momentum tensor itself \cite{Bakker:2004ib,Leader:2011cr,Ji:2012sj}
\beq
J^{\mu\nu} = \int d^3x M^{+\mu\nu}\,, \label{angular}
\eeq
 where $d^3x = dx^- d^2x_\perp$ in the light--front form. The other is the Pauli--Lubanski vector \cite{Harindranath:1999ve,Ji:2012vj}
\beq
W^\mu = -\frac{1}{2}\epsilon^\mu_{\ \, \nu\rho\sigma}P^\nu  \int d^3x M^{+\rho\sigma}\,.
 \eeq
% \beq
% M^{\lambda \mu\nu} = x^\mu T^{\lambda \nu} -x^\nu T^{\lambda \mu}\,, \label{where}
% \eeq
% $M^{\lambda\mu\nu}$ is the angular momentum tensor.
  %which  the energy--momentum tensor $T^{\mu\nu}$.
  In the longitudinally polarized case, the distinction is irrelevant because only one component of $M$ is involved
  \beq
  W^+  = \frac{1}{2}\epsilon^{ij}P^+\int d^3x M^+_{\ \, ij} \propto J^{12}\,.
  \eeq
However, in the transversely polarized case it is expected on general grounds that any transverse spin sum rule based on $\langle J^{\mu\nu}\rangle $ has frame--dependence \cite{Leader:2011cr,Leader:2012md}. Following \cite{Ji:2012vj}, in this paper we only consider the expectation value of the Pauli--Lubanski vector $\langle W^i\rangle$ where
 \beq
 W^i = \epsilon^{ij}\left(P^- \int d^3x M^{++}_{\quad \ j} -P^+\int d^3x M^{+-}_{\ \ \ \ j} \right)\,, \label{parton}
\eeq
in pursuit of a frame--independent  sum rule.

  According to (\ref{parton}), the partonic decomposition of $W^i$ boils down to that of $M^{++i}$ and $M^{+-i}$. The latter is apparently of higher twist,
%twist--four,
but we shall shortly see that it nevertheless gives equally important contributions as the $M^{++i}$ term.
In fact, the helicity contributions of quarks and gluons entirely come from $M^{+-i}$.
In the quark case, we have
\beq
 M_{q-spin}^{\mu\nu\lambda} = -\frac{1}{2}\epsilon^{\mu\nu\lambda\sigma}\bar{\psi}
\gamma_5 \gamma_\sigma  \psi\,,  \label{qspin}
\eeq
so that $M_{q-spin}^{++i}=0$ and
\beq
M_{q-spin}^{+-i} = \frac{1}{2}\epsilon^{ij}\bar{\psi} \gamma_5\gamma_j\psi\,.
\eeq
This immediately gives the usual quark helicity contribution $\Delta \Sigma$
 \beq
\frac{\langle PS|W^i_{q-spin}|PS\rangle}{2P^+ (2\pi)^3\delta^3(0)}
 = \frac{1}{4} \frac{\langle PS| \int d^3x \bar{\psi}\gamma_5\gamma^i\psi |PS\rangle}{(2\pi)^3\delta^3(0)} = \frac{1}{2}\int dx g_T(x) S^i = \frac{1}{2}\Delta \Sigma S^i\,,
 \label{deltaQ}
 \eeq
  where $g_T(x)$ is defined in (\ref{gt}).
We thus concur with \cite{Ji:2012vj} that the quark helicity relevant to the nucleon spin sum rule
is  the first moment of the twist--three distribution $g_T(x)$,
and not  that of the twist--two transversity distribution $h_1(x)$ \cite{Bakker:2004ib}.
This is indeed a direct consequence of rotational invariance in the rest frame of the nucleon.
%and suggests that the transverse-spin {\it average} $g_T(x)$ is relevant to the transverse spin sum rule, i.e.,
%the decomposition of the transverse spin.
%One may expect a similar situation for the gluonic contribution to the nucleon spin.

In the gluon case we use the recently developed gauge invariant framework of spin decomposition in which the gluon helicity component reads \cite{Chen:2008ag,Wakamatsu:2010cb,Hatta:2011zs}
\beq
 M_{g-spin}^{\mu\nu\lambda}&=&  F^{\mu\lambda}A_{phys}^{\nu } -
F^{\mu\nu}A_{phys}^{\lambda }  \,,  \label{glu}
\eeq
 where $A_{phys}^\mu$ is the `physical part' of the gauge field \cite{Hatta:2011zs}
 \beq
 A_{phys}^\alpha (\lambda n) = -\int d\zeta {\mathcal K}(\zeta-\lambda)W_{\lambda \zeta} \, n_\mu F^{\mu\alpha}(\zeta n)\,,  \label{aphys}
 \eeq
 with ${\mathcal K}(\lambda)$ being either $\frac{1}{2}\epsilon(\lambda)$, or $\pm \theta (\pm \lambda)$.
The matrix element of (\ref{glu}) is related to $\Delta G$, both for the transverse and longitudinal polarizations.   Indeed,
dividing (\ref{1}) by $x$ and integrating over $x$,\footnote{When integrating over $x$, one has to specify the $i\epsilon$--prescription for the pole $1/x$. Different prescriptions correspond to the different kernels ${\mathcal K}$ in (\ref{aphys}). } we find, using $\int_{-1}^1 dx\,   G(x)=0$,
\beq
\langle PS | F^{+\alpha}(0) A^\beta_{phys}(0) |PS\rangle &=&
 \frac{1}{2}\epsilon^{+-\alpha\beta} S^+ \int dx \Delta G(x)+\epsilon^{+\alpha\beta j}S_j \int dx {\cal G}_{3T}(x)
 \nonumber \\ &=& \epsilon^{+\alpha\beta \mu}S_\mu \Delta G \,.   \label{hen}
 \eeq
 As seen in (\ref{hen}),  the matrix element of operators involving $A_{phys}^\mu(0)$,  despite being nonlocal,  does not depend on the noncovariant vector $n^\mu$. This is because the scale transformation $n^\mu \to cn^\mu$ applied to (\ref{aphys}) with $\lambda=0$ can be absorbed by the rescaling  $\zeta \to \zeta/c$ of the integration variable.
Noticing that  $A_{phys}^+=0$ and therefore
\beq
  M_{g-spin}^{++i}=0\,, \qquad
 M_{g-spin}^{+-i}&=&  F^{+i}A_{phys}^{-} -
F^{+-}A_{phys}^{i }  \,,
\eeq
 we obtain
\beq
\frac{\langle PS|W^i_{g-spin}|PS\rangle}{2P^+ (2\pi)^3\delta^3(0)}  = S^i \int dx\, {\cal G}_{3T}(x) =S^i \Delta G\,. \label{deltaG}
\eeq
We thus see that $\Delta G$ is the gluon helicity contribution also for the transverse polarization,
due to the contribution of the Wandzura--Wilczek part of ${\cal G}_{3T}(x)$ as fully given by (\ref{solve}).
%and ${\cal G}_{3T}(x)$ as fully given by (\ref{solve}) can be understood as its partonic density.

\subsection{Complete transverse spin decomposition?}

 Now that we have seen the helicity components, (\ref{qspin}) and (\ref{glu}), of the decomposition scheme  \cite{Chen:2008ag,Hatta:2011zs} works also for the transverse polarization, it seems a straightforward task to demonstrate the complete decomposition of the transverse spin
 including the \emph{canonical} orbital angular momentum of quarks and gluons
 \beq
 M^{\mu\nu\lambda}_{q-orbit} =\bar{\psi} \gamma^\mu (x^\nu iD_{pure}^\lambda -x^\lambda iD_{pure}^\nu)\psi \,, \quad M^{\mu\nu\lambda}_{g-orbit} =F^{\mu\alpha}(x^\lambda D^\nu_{pure} - x^\nu D^\lambda_{pure})A^{phys}_\alpha \,,  \label{oam}
 \eeq
  where $D^\mu_{pure}=D^\mu-igA^\mu_{phys}$, as has been done  in the longitudinally polarized case \cite{Hatta:2012cs}. In doing so, however, we encountered an unexpected difficulty which actually already arises in the more conventional decomposition scheme by Ji \cite{Ji:1996ek} recently revisited   in \cite{Ji:2012vj}. In Ji's scheme, the angular momentum tensor  is given by
 \beq
 M_{q,g}^{\lambda \mu\nu} = x^\mu T_{q,g}^{\lambda \nu} -x^\nu T_{q,g}^{\lambda \mu}\,, \label{where}
 \eeq
  where $T_{q,g}$ is the (Belinfante--improved) energy momentum tensor of quarks/gluons whose matrix element is   parameterized as
 \beq
 \langle P'S'|T_{q,g}^{\mu\nu}|PS\rangle &=& \bar{u}(P'S')\Bigl[A_{q,g} \gamma^{(\mu}\bar{P}^{\nu)}
 +B_{q,g}\frac{\bar{P}^{(\mu}i\sigma^{\nu)\alpha}\Delta_\alpha}{2M} \nonumber \\
 && \quad  + C_{q,g}\frac{\Delta^\mu\Delta^\nu -g^{\mu\nu}\Delta^2}{M} + \bar{C}_{q,g}Mg^{\mu\nu}\Bigr] u(PS)\,,  \label{para}
 \eeq
 where $\bar{P}^\mu=\frac{1}{2}(P^\mu+P'^\mu)$, $\Delta^\mu = P'^\mu-P^\mu$.
  Ref.~\cite{Ji:2012vj} concludes that the sum rule based on the Pauli--Lubanski vector is nothing but the Ji sum rule
\beq
\frac{\langle PS|W^i_{q,g}|PS\rangle}{2P^+ (2\pi)^3\delta^3(0)}  \equiv J_{q,g} S^i= \frac{1}{2}(A_{q,g}+B_{q,g})S^i\,, \label{pl}
\eeq
 where $J_q+J_g=\frac{1}{2}$ is the total nucleon spin.

Although (\ref{pl}) seems appealing, as its frame--independent form exemplifies the advantage of using the Pauli--Lubanski vector, it should be corrected as follows:
%as in (\ref{order}) below
As noted already, the transverse angular momentum receives a contribution from the twist--four matrix element $\langle M^{+-i}_{q,g}\rangle \sim \langle x^iT^{+-}_{q,g}\rangle \sim i\frac{\partial}{\partial \Delta_i} \langle T^{+-}_{q,g}\rangle$ which involves the derivative with respect to the momentum transfer $\Delta$. The problem is that the nonforward spinor product
\beq
\bar{u}(P'S') u(PS){\mathop{\approx}_{\Delta^\mu \to 0}} 2M +i\frac{ \bar{P}^3 }{M(\bar{P}^0 +M)}\epsilon^{ij}\Delta_i S_j\,,
\eeq
contains an order ${\mathcal O}(\Delta)$, and manifestly frame--dependent (velocity--dependent)  term in the transversely polarized case, as first observed in \cite{Bakker:2004ib}.
There are two sources of $\bar{u}(P'S')u(PS)$  in (\ref{para}). The first term of $\langle T^{+-}_{q,g}\rangle$ (proportional to $A_{q,g}$) contains it as one can see from the Gordon identity
\beq
\bar{u}(P'S')\gamma^\mu u(PS) &=& \frac{\bar{P}^\mu}{M}\bar{u}(P'S')u(PS) +\bar{u}(P'S')\frac{i\sigma^{\mu\alpha}\Delta_\alpha }{2M}u(PS)\,.
\label{ubar}
\eeq
Fortunately, this unwanted term cancels against a similar term from $\langle T^{++}_{q,g}\rangle$ and does not contribute to $\langle W^i_{q,g}\rangle$ \cite{Ji:2012vj}. However, the fourth term
\beq
\bar{C}_{q,g} Mg^{+-}\bar{u}(P'S')u(PS)\,, \label{ano}
\eeq
which is related to the QCD trace anomaly and has  no compensating term from $\langle T^{++}_{q,g}\rangle$, leads to an additional frame--dependent contribution to $J_{q,g}$
 \beq
 J_{q,g}=\frac{1}{2}(A_{q,g}+B_{q,g}) +  \frac{P^3}{2(P^0+M)}\bar{C}_{q,g}\,, \label{order}
 \eeq
   which is parametrically unsuppressed if $\bar{C}_{q,g} \sim {\mathcal O}(1)$  and  the motion is relativistic $P^3 \sim{\mathcal O}(P^0)$. It should be noted that the sum vanishes  $\bar{C}_q+\bar{C}_g=0$ due to the conservation of the total energy momentum tensor $T^{\mu\nu}=T^{\mu\nu}_q + T^{\mu\nu}_g$,  so they cancel in the \emph{total}  angular momentum $\frac{1}{2}=J_q+J_g$.  However, they do affect the decomposition \emph{into} the quark and gluon contributions.\footnote{Incidentally, we note that the extra term in (\ref{order}) is unrelated to the frame--dependent term asserted in \cite{Leader:2011cr} which grows unboundedly with increasing energy.}

The above  problem fortunately does not arise in the longitudinally polarized case where the product $\bar{u}'u$ does not contain an ${\mathcal O}(\Delta)$ term. Both the Ji decomposition and the complete decomposition \cite{Hatta:2012cs} stand unaffected in this case. However, in the transversely polarized case the Ji decomposition is afflicted by the (nominally) twist--four  non-covariant terms, and the same problem seems to persist in the expectation value of the canonical orbital angular momenta $\langle W^i_{q-orbit}\rangle$ and $\langle W^i_{g-orbit}\rangle$ computed from (\ref{oam})
\cite{prep}.  This fact, together with (\ref{deltaQ}) and (\ref{deltaG}), implies that in the transversely polarized case we may only achieve the decomposition
 \beq
 \frac{1}{2} = \frac{1}{2}\Delta \Sigma + \Delta G + L\,,
 \eeq
 where the canonical orbital angular momentum $L$
  cannot be separated into those of quarks and gluons in a  frame--independent way.

%\section{5. Conclusions}

\section*{Acknowledgements}
We thank Feng Yuan for discussions and for sending us a copy of \cite{Ji:2012vj} before it became publicly available.
 We also thank Elliot Leader for discussions.
The work of K.~T. is supported in part by the Grant-in-Aid for Scientific
Research (Nos.~23540292 and 24540284). The work of S.~Y. is supported by the Grant-in-Aid for Scientific
Research (No. 21340049).\\

\emph{Note added:} A few days after the submission of this work on arXiv, a preprint \cite{Leader:2012ar} appeared which pointed out essentially the same non-covariant term as in (\ref{order}).

\end{document}